\documentclass[useAMS,usenatbib]{mn2e}

\usepackage[usenames,dvips]{color}
\usepackage{amsmath}
\usepackage{graphicx}
\usepackage{defs}

\title[Fingerprints of the ICs]{Fingerprints of the initial conditions on the density profiles of cold and warm dark matter haloes}

\author[E. Polisensky and M. Ricotti]{E. Polisensky$^{1}$\thanks{E-mail: Emil.Polisensky@nrl.navy.mil} and M. Ricotti$^{2}$\footnotemark[1]\thanks{E-mail: ricotti@astro.umd.edu}\\
$^{1}$Naval Research Laboratory, Washington, D.C. 20375, USA\\
$^{2}$Department of Astronomy, University of Maryland, College Park, Maryland 20745, USA}
\begin{document}

\date{\today}

\pagerange{\pageref{firstpage}--\pageref{lastpage}} \pubyear{2014}

\maketitle

\label{firstpage}

\begin{abstract}

We use N-body simulations of dark matter haloes in cold dark matter (CDM) 
and a large set of different warm dark matter (WDM) cosmologies to 
demonstrate that the spherically averaged density profile of dark matter 
haloes has a shape that depends on the power spectrum of matter 
perturbations. Density profiles are steeper in WDM but become shallower 
at $r<0.01 R_{vir}$. Virialization isotropizes the velocity dispersion 
in the inner regions of the halo but does not erase the memory of the 
initial conditions in phase space. 
The location of the observed deviations from CDM in the density profile and in phase space can be directly related to the ratio between the halo mass and the filtering mass and are most evident in small mass haloes, even for a 34~keV thermal relic WDM. The rearrangement of mass within the haloes supports analytic models of halo structure that include angular momentum. We also find evidence of a dependence of the slope of the inner density profile in CDM cosmologies on the halo mass with more massive haloes exhibiting steeper profiles, in agreement with the model predictions and with previous simulation results. Our work complements recent studies of microhaloes near the filtering scale in CDM and strongly argue against a universal shape for the density profile.

\end{abstract}


\section{Introduction}\label{sec:1}

The seminal work of \citet{nav1996} found that the density structure 
of relaxed dark matter haloes are well represented by what has become 
known as the NFW profile:
\begin{equation}\label{eqNFW_4}
\rho(r) = \frac{\rho_{s}}{(r/r_s)(1+r/r_s)^2},\\
\end{equation}
where $\rho(r)$ is the density in a spherical shell at distance $r$ 
from the halo centre. By scaling the free parameters $r_s$ and $\rho_s$, 
which define a characteristic length and density, the NFW profile can 
describe dark matter haloes from dwarf galaxy to cluster scales.
Furthermore, it was found the NFW profile was valid for haloes 
regardless not only of mass but also the power spectrum of initial 
density fluctuations and values of cosmological parameters, establishing 
that density profiles are universal in form independent of the 
cosmological context \citep{nav1997}. Another universal property was 
found in the coarse-grained phase-space density profile, 
$Q \equiv \rho/\sigma^3$, where $\sigma$ is the velocity dispersion of 
simulation particles. \citet{tay2001} discovered $Q$ has a remarkably 
simple form of a power-law, $Q \propto r^\gamma$, with $\gamma \sim -1.9$.

It is useful to recast the free parameters of the NFW profile in terms of a 
halo mass and concentration. For relaxed haloes a radius can be defined 
in which the material has reached virial equilibrium:
\begin{equation}\label{eqnMvir}
M_{vir} = \frac{4\pi}{3} \Delta(z) \rho_{c}(z) R_{vir}^3,\\
\end{equation}
where $R_{vir}$ is the virial radius enclosing a density $\Delta$ times the 
critical density, $\rho_{c}$, at redshift $z$. The characteristic 
radius $r_{s}$ can be recast as the concentration parameter, 
$c_{vir} \equiv R_{vir}/r_s$. Much effort has gone into understanding 
the relationship between $c_{vir}$ and $M_{vir}$ as well as the dependencies 
on the background cosmology and the evolution with redshift \citep{pra2012}. 
This is necessary for predicting the properties of luminous galaxies 
that reside in the dark matter haloes and for using galaxy observations 
as probes of the cold dark matter (CDM) paradigm. The concentration was found to correlate 
with mass such that smaller mass haloes are more concentrated \citep{nav1997}.
This was understood as a consequence of the earlier formation epoch of 
small mass haloes in the bottom-up structure formation of CDM.
Since small haloes collapse earlier, their inner regions reflect the 
higher universal density of matter at earlier times.
Changing the cosmological parameters or the power spectrum changes 
the halo formation epoch and affects the concentrations but does not affect
the shape of the universal profile \citep{pol2014}.
This interpretation is consistent with simulations of hot and warm dark 
matter which found haloes with masses below the truncation scale 
form later and have lower concentrations than CDM haloes of similar 
size \citep{avi2001,bod2001,kne2002}.

Much effort has also gone into understanding the physical processes 
that produce the NFW profile. There are two main approaches to 
analytically modeling the density profile: smooth accretion based on 
spherical infall \citep{gun1972,got1975}, and hierarchical merging following 
Press-Schechter formalism \citep{pre1974,pee1974,lac1993,man2003}.
Both approaches have been successful at producing the universal profiles. 
This has been explained as a result of the process of virialization. 
If virialization erases all information about the past merging history 
of the halo then it does not matter if the mass accretion is modeled as 
clumpy or smooth. However, a consensus has not emerged on the dominant 
processes occurring during virialization or if the virialization process 
erases all memory of the initial conditions.

The early stages of halo formation are marked by rapid accretion and 
mergers making it natural to consider violent relaxation as the dominant 
mechanism determining the dark matter profiles in the fluctuating 
gravitational potential \citep{whi1996b}. Violent relaxation was originally 
proposed to explain the structure of elliptical galaxies \citep{lyn1967} 
where estimates of star-star encounters would not establish equilibrium 
in a Hubble time. The relaxation time of a forming halo is related 
to the rate of change of the gravitational potential.
\citet{aus2005} and \citet{bar2006} argue the universal nature of 
$Q(r)$ results from violent relaxation.

The works of \citet{wec2002,zha2003b,zha2009} have shown there are two main 
eras of halo growth, a fast accretion phase and a slow phase. 
The fast growth phase in CDM is dominated by mergers of objects with 
similar mass in contrast to the slow growth phase characterized by 
quiescent accretion and minor mergers. The inner halo is set at the end 
of the fast era with the slow growth phase having little impact on the 
inner structure and gravitational potential well, leading to an 
inside-out growth of haloes. These studies find violent relaxation is 
only important in forming the inner profile with the outer profile 
determined by secondary infall during the slow growth phase.

The NFW profile is characterized by a logarithmic slope, 
$\alpha \equiv d \log \rho/d \log r$, that rolls from an asymptotic 
value $\alpha = -3$ at large radii to $\alpha = -1$ in the inner halo.
The value of the inner slope has been a matter of controversy. 
The first concerned the value of the asymptotic slope \citep{moo1999b}.
As the number of particles in simulations have increased it has become evident 
the density profiles do not approach an asymptotic value 
but continue to roll slowly with radius \citep{nav2004,die2004,gra2006} 
and are better described by Einasto profiles \citep{ein1965}. 
However, this has not changed the conclusion that all information about 
the formation history is lost in the virialization process.

The second controversy is a dependence of the inner profile on the halo mass. 
Many models have been constructed that explain the emergence of 
the universal profile as a consequence of repeated mergers 
\citep{sye1998,nus1999,sub2000,dek2003}. Although they differ in 
the details, the relevant physical processes determining the halo 
properties are the tidal stripping of material from accreting subhaloes, 
dynamical friction, and tidal compression transferring energy from the 
satellites to the halo particles and the decaying of satellite orbits 
to the halo centre. These models predict a dependence of the inner density 
profile slope on the slope of the power spectrum at the scale of the 
halo, $P(k) \propto k^n$. The steeper spectrum characteristic of 
dwarf-scales is predicted to produce softer cores than for galactic 
and cluster-scale haloes. Independent of the merger models, 
\citet{del2010} questions the universality of both the density 
and the $Q$ profile and concludes both should depend on mass.
His spherical infall models with angular momentum show a steepening of 
the inner density profile with increasing halo mass, although to a lesser 
extent than the merger models. The central question of these studies is, 
do haloes in equilibrium retain any memory of the initial conditions and 
mass function of accreting satellites they are built from or is all 
information lost in the virialization process?

\citet{ric2003} ran CDM simulations of the same realization of the density 
field in boxes of varying side length to compare the profiles at different 
mass scales. He examined the average profiles when the box structures showed 
similar clustering and the most massive haloes were composed of the 
same number of particles. He found a systematic dependence of the inner 
slope on halo mass with dwarf-scale haloes having softer cores than 
galactic and cluster-scale haloes in agreement with the predictions of 
\citet{sub2000}. These results were reinforced in \citet{ric2004a,ric2007}.
\citet{jin2000} also saw a dependence of inner slope in their simulations 
of haloes at galactic and cluster scales.

Another way of testing the importance of substructure is by introducing 
a truncation in the power spectrum as in hot and warm dark matter (WDM)
cosmologies where substructure is suppressed below the particle 
free-streaming scale and haloes form by monolithic collapse.
Many investigations using these cosmologies have been conducted 
(e.g. \citealt{wan2009}, \citealt{hus1999a}, \citealt{moo1999}, 
\citealt{col2000}, \citealt{bus2007}, \citealt{bod2001}, \citealt{pol2011}). 
These works find haloes that form below the truncation mass have lower 
concentrations consistent with their later formation epochs 
but the profiles are well described by the NFW form. This is in contrast 
to \citet{col2008} who find the inner profiles are systematically 
steeper in their WDM simulations of Galaxy-sized haloes. Recent studies of 
CDM microhaloes confirm halo profiles near the streaming scale are steeper 
when a truncation is introduced into the power spectrum than without, however 
the origin of the steepening remains unclear \citep{ish2010,and2013b,ish2014}. 
Independent models of the density profiles 
of WDM haloes predict not a steepening but a flattening of the inner profile 
due to the truncated power spectrum \citep{wil2004,vin2012}. 
It is important to stress the flattening in these models is due to the 
truncated power spectrum alone, not to the random thermal motions of WDM 
particles which were not included in the models.

In this work we employ N-body simulations of halo formation in CDM and
WDM cosmologies to explore the effects of the power spectrum on halo
structure and dynamics. We use the method of \citet{ric2003} of
scaling the simulation volume to change the mass scale
(\S~\ref{method}). However, we do not study a statistical sample
of haloes but focus on a handful of halos with about a factor of 100 greater
mass resolution than in \citet{ric2003}. The goal is not to rigorously
test any particular halo model but simply to look for evidence the
halo retains memory of the initial power spectrum. This evidence is
expected to manifest itself as trends in the halo profiles as the mass
and truncation scales change (\S~\ref{results}). We examine if our
results are typical of a larger halo population (\S~\ref{secVAR}) and
investigate the physical origin of the results (\S~\ref{origin}). A
discussion and comparison to previous published works is presented in
\S~\ref{disc} and the summary in \S~\ref{summary}.

\section{Numerical Simulations}\label{method}

\subsection{Cosmological Models}

WDM particles are relativistic in the early universe and free-stream out 
of overdense regions before the adiabatic expansion of the 
universe reduces the particles to subrelativistic velocities. 
WDM thus damps density perturbations below a characteristic scale that depends 
on the particle mass and acts as a filter on the power spectrum of density 
perturbations. The power spectra for WDM cosmologies is related to that 
for CDM by
\begin{equation}\label{eq1_4}
P_{W}(k) = P_{C} T_{W}^2,\\
\end{equation}
where $T_{W}$ is the WDM transfer function. The transfer function given 
by \citet{bod2001} is used for dark matter particles that were coupled 
to the relativistic cosmic plasma at early times and achieved thermal 
equilibrium prior to the time of their decoupling. The formula of 
\citet{eis1997} is adopted for the CDM power spectrum.

We define the WDM filtering mass as in \citet{som2001},
\begin{equation}
M_f \equiv \frac{4\pi^4}{3} \Omega_m \rho_c k_f^{-3},\\
\end{equation}
where $\rho_c$ is the critical density and $k_f$ is a characteristic 
free-streaming wave number defined where $T_{W}^2=0.5$. For consistency 
with \citet{som2001} we also define the free-streaming, or filtering 
length as $R_f \equiv 0.46 k_f^{-1}$.

We adopted values for cosmological parameters from the Bolshoi 
simulation \citep{kly2010}, 
($\Omega_m$, $\Omega_{\Lambda}$, $\Omega_b$, $h$, $\sigma_8$, $n_s$) = 
(0.27, 0.73, 0.0469, 0.7, 0.82, 0.95),
which were chosen to be within $1\sigma$ of WMAP5, WMAP7, and consistent 
with the results of supernovae, and X-ray cluster surveys. These parameters 
are also within $1.7\sigma$ of WMAP9 and $2.2\sigma$ of Planck1.
We use a variety of WDM models for thermal relics in the range 
$0.75-66$~keV.

Since our focus is to examine the effects of the power spectrum on halo 
structure, the initial conditions include particle velocities due to the 
gravitational potential using the Zeldovich approximation but random thermal 
velocities appropriate for WDM have not been added to the simulation particles.
For the WDM cosmologies adopted here the effects of thermal velocities are 
expected to be small; this is discussed further in Section~\ref{disc}.

\subsection{Software}

The simulations were conducted with the $N$-body cosmological simulation 
code {\sevensize GADGET-2} \citep{spr2005} with gravitational physics 
only and initial conditions generated with the {\sevensize GRAFIC2} software 
package \citep{ber2001}. We produce a single realization of the density 
field but vary the power spectrum of fluctuations appropriate for CDM 
and WDM cosmologies.

The AMIGA's Halo Finder ({\sevensize AHF}) software \citep{kno2009} 
was used to identify and characterize all gravitationally bound 
haloes composed of at least 50 particles after iteratively 
removing unbound particles. The virial mass of a halo is defined 
in Equation~\ref{eqnMvir}. Since the simulations are confined to high 
redshifts ($z>3$) the universe is matter dominated at all epochs 
and we adopt the virial condition for an Einstein-de Sitter 
cosmology, $\Delta(z)=178$. The {\sevensize MergerTree} tool 
in {\sevensize AHF} was used to construct merger trees, identify halo 
progenitors at all times, and for identifying haloes across cosmologies.

{\sevensize AHF} calculates the convergence radius according to the criterion 
of \citet{pow2003} and is generally about 10 softening lengths, 
enclosing $\sim 2000$ particles at $r\sim 0.006 R_{vir}$. 
We tested this by running low resolution simulations 
and found that the profiles are actually converged to about $5-6$ 
softening lengths, enclosing $\sim 200$ particles at $r\sim 0.003 R_{vir}$. 
The convergence radius given by {\sevensize AHF} may be overly conservative 
for the simulations but this has no impact on the results. When examining 
the halo profiles we adopt the convention of plotting $r>6\epsilon$ but 
indicating in bold where the profiles satisfy the criterion 
of \citeauthor{pow2003}

\subsection{Simulations}

We simulated a small cubic box with a comoving side length of 3.3~Mpc 
from $z=79$ to $z=8$ with $512^3$ particles and mass resolution 
$\sim 10^4 M_{\sun}$. A halo of mass $\sim 2 \times 10^8 M_{\sun}$ that 
appeared to have an early formation epoch and relaxed to virial 
equilibrium at scale factor $a=0.1$ ($z \equiv a^{-1} -1$) 
was chosen for resimulation using a zoom technique. We refer to this 
as ``Halo A.'' A volume of higher mass resolution was generated 
in the initial conditions covering the initial volume of particles 
that end within three virial radii of Halo A. We ran high resolution 
simulations with a mass resolution of 18.7~$M_{\odot}$ in CDM and multiple 
WDM cosmologies in the range $4-66$~keV. We adopted a force softening 
length of $8$~pc, held constant in comoving units. To test the convergence 
of our results we also ran low resolution simulations with the mass 
resolution reduced by a factor of $8$ in CDM and 6~keV WDM. We further 
tested the dependence of our results on the initial conditions by running 
low resolution tests in CDM and 6~keV starting from $z=120$.

\begin{figure}
\begin{center}
\includegraphics*[height=0.9\columnwidth,angle=270]{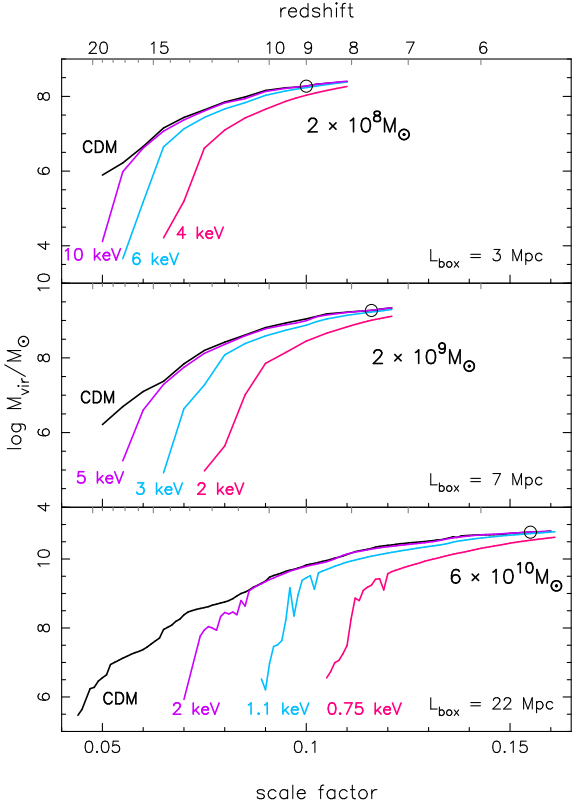}
\caption{Mass growth of Halo A in the small, medium, and large box simulations 
(top to bottom) in CDM and select WDM cosmologies. The circles show the 
CDM halo at the normalization times when the halo has grown to 
$\sim 10^7$ particles at the three mass scales.\label{figMASSA}}
\end{center}
\end{figure}

We ran additional simulations with the box size increased to medium and 
large side lengths of 7.0 and 22.4~Mpc to increase the mass scale a 
factor of 10 and 320, respectively. The force softening lengths were also 
scaled with the box size to 17~pc and 55~pc, respectively. CDM and WDM 
cosmologies ranging from $2-5$~keV were run for the medium mass scale while 
CDM and $0.75-2$~keV WDM were run for the large mass scale. To compare 
Halo A across mass scales we define ``normalization times'' as the epochs 
when the CDM haloes have grown to encompass the same number of particles 
within the virial radius, $N \sim 10^7$, as the small mass scale at $a=0.1$. 
This occurred at $a=0.116$ and $a=0.155$ for the medium and large scales, 
respectively. Figure~\ref{figMASSA} shows the growth of Halo A for the three 
mass scales. Halo formation is delayed in the WDM cosmologies. However, once 
it begins it grows quickly until it catches up with the CDM halo, after which 
it evolves at a similar rate. The circles show the normalization times when 
the halo has entered the slow-growth phase and has reached the same size 
in CDM at all scales. At the normalization times the halo masses are 
approximately $2 \times 10^8 M_{\sun}$, $2 \times 10^9 M_{\sun}$, and 
$6 \times 10^{10} M_{\sun}$ for the small, medium, and large mass scales, 
respectively.

Table~\ref{tab1} gives a summary of the simulations conducted in this work.
Listed in the table are the WDM filtering mass, the filtering length, 
and the number of simulation particles sampling the filtering mass, $N_f$.
For the WDM cosmologies the box side length $L_{box}$ is given in units of the 
filtering length. This is a convenient way to show in which simulations 
the effects of the truncated power spectrum will be similar across mass scales.
The 2~keV simulation in the large box is expected to be similar to 
the 5~keV medium box and 10~keV small box. Likewise, the 1.1~keV large box 
will be similar to the 3~keV medium box and 6~keV small box. The 0.75~keV 
large box will be similar to the 2~keV medium box and 4~keV small box.
These simulations are colour-coded in Figure~\ref{figMASSA} according 
to their ratios of box scale to WDM filtering scale. 
Another way of characterizing the similarity of these simulations is by the 
ratio of filtering mass to halo virial mass. 
This ratio is equivalent to $N_f$ 
expressed in units of $10^7$, as listed in column $(4)$ of Table~\ref{tab1}. 
For the similar cosmologies given above, the filtering masses are 
approximately $7\%$, $40\%$, and $170\%$ of the halo masses at the 
normalization times.

\begin{table}
\caption{Properties of simulations.\label{tab1}}
\begin{center}
\begin{tabular}{l c c c c}\hline
 Cosmo & $M_{f}$ & $R_f$ & $N_f$ & $L_{box}$ \\
 & [$M_{\odot}$] & [kpc] & [$\times 10^7$] & \\
(1) & (2) & (3) & (4) & (5)\\
\hline
\hline
\multicolumn{5}{c}{\textit{Small Box: $m_{res}=18.7$ $M_{\odot}$}}\\
\hline
\hline
 CDM & - & - & - & 3270 kpc \\
 66 keV & $1.90\times10^4$ & 0.7 & $10^{-4}$ & 4485 $R_f$ \\
 34 keV & $1.90\times10^5$ & 1.6 & $0.001$ & 2082 $R_f$ \\
 21 keV & $9.50\times10^5$ & 2.7 & $0.005$ & 1217 $R_f$ \\
 15 keV & $3.19\times10^6$ & 4.0 & $0.02$ & 812 $R_f$ \\
 10 keV & $1.29\times10^7$ & 6.4 & $0.07$ & 510 $R_f$ \\
 7 keV & $4.43\times10^7$ & 9.7 & $0.2$ & 338 $R_f$ \\
 6 keV & $7.54\times10^7$ & 11.5 & $0.4$ & 283 $R_f$ \\
 5 keV & $1.41\times10^8$ & 14.2 & $0.8$ & 230 $R_f$ \\
 4.5 keV & $2.03\times10^8$ & 16.1 & $1.1$ & 203 $R_f$ \\
 4 keV & $3.05\times10^8$ & 18.4 & $1.6$ & 178 $R_f$ \\
\hline
\hline
\multicolumn{5}{c}{\textit{Medium Box: $m_{res}=187$ $M_{\odot}$}}\\
\hline
\hline
 CDM & - & - & - & 7049 kpc \\
 5 keV & $1.41\times10^8$ & 14.2 & $0.08$ & 495 $R_f$ \\
 3 keV & $8.24\times10^8$ & 25.6 & $0.4$ & 275 $R_f$ \\
 2 keV & $3.34\times10^9$ & 40.8 & $1.8$ & 173 $R_f$ \\
\hline
\hline
\multicolumn{5}{c}{\textit{Large Box: $m_{res}=6,010$ $M_{\odot}$}}\\
\hline
\hline
 CDM & - & - & - & 22394 kpc \\
 2 keV & $3.34\times10^9$ & 40.8 & $0.06$ & 548 $R_f$ \\
 1.1 keV & $2.63\times10^{10}$ & 81.2 & $0.4$ & 276 $R_f$ \\
 0.75 keV & $9.84\times10^{10}$ & 126.2 & $1.6$ & 177 $R_f$ \\
\hline
\end{tabular}
\end{center}
\end{table}

Table~\ref{tab2} summarizes the properties of Halo A at the normalization 
times. An examination of Figure~\ref{figMASSA} shows Halo A has not 
suffered a recent major merger and is in the slow growth phase in all 
cosmologies at the normalization times. However, a more rigourous 
examination of the halo relaxation state is desirable. 
Differences from a universal profile are seen in unrelaxed 
haloes and haloes with large amounts of substructure \citep{jin2000a}.
Additionally, the inner slope of the density profile is sensitive to the 
location of the halo centre. An artificial flattening of the profile could 
be produced by an ambiguously defined centre due to a recently arrived 
subhalo at the core, for example. We performed a qualitative visual 
examination that the halo centres determined by {\sevensize AHF} 
correspond to the density peak of particles and we examined quantitative 
measures of the relaxation. Studies with large samples of haloes have 
identified several metrics for separating haloes by relaxation state 
\citep{net2007,mac2007,mac2008}:  
$x_{off}$, the offset between the halo centre and centre of mass of 
particles within $R_{vir}$; the virial ratio $2K/|U| - 1$; the mass fraction 
bound in subhaloes $f_{sub}$; and the spin parameter, $\lambda^{\prime}$, 
that characterizes the halo angular momentum \citep{bul2001}:
\begin{equation}\label{lamb}
  \lambda^{\prime} = \frac{J}{\sqrt{2}M_{vir}v_{vir}R_{vir}},\\
\end{equation}
where $J$ is the total angular momentum of all particles within 
$R_{vir}$ and $v_{vir}$ is the circular velocity at $R_{vir}$, $v^2 \equiv GM/R$.
These metrics are listed in Table~\ref{tab2}. 
The general conditions for a relaxed halo are: 
$\lambda^{\prime} < 0.1$, $x_{off} < 0.07 R_{vir}$, $2K/|U|-1 < 0.35$, 
and $f_{sub} < 0.1$ \citep{lud2012,lud2013,lud2014}. Halo A largely satisfies 
these criteria with the greatest discrepancy being a slightly larger virial 
ratio of $\sim 0.4$, however in Section~\ref{secVAR} we examine the stability 
of the profiles and find our results are not due to the relaxation state 
of the halo or transient accretion events.

We note a curious increase in $f_{sub}$ for the warmest cosmologies in all 
three box sizes. We examined this further by applying varying subhalo mass 
cuts and found $f_{sub}$ is dominated by subhaloes with less than 5000 
particles in these cosmologies. Artifical haloes of this size are known to 
form along filaments in truncated power spectrum cosmologies \citep{wan2007} 
and are likely contaminating $f_{sub}$.

\begin{table}
\caption{Properties of Halo A at the normalization times in the high resolution simulations.\label{tab2}}
\begin{center}
\begin{tabular}{l c c c c c}\hline
 Cosmo & $M_{vir}$ & $\lambda^{\prime}$ & $x_{off}$ & $\frac{2K}{|U|} - 1$ & $f_{sub}$ \\
 & [$10^{8}M_{\odot}$] & [$10^{-2}$] & [$R_{vir}$] & & \\
(1) & (2) & (3) & (4) & (5) & (6) \\
\hline
\hline
\multicolumn{6}{c}{\textit{Halo A - Small Box}}\\
\hline
\hline
 CDM & 1.868 & 4.21 & 0.06 & 0.41 & 0.08\\
 66 keV & 1.847 & 3.84 & 0.05 & 0.40 & 0.07\\
 34 keV & 1.863 & 3.89 & 0.05 & 0.40 & 0.06\\
 21 keV & 1.882 & 3.99 & 0.06 & 0.39 & 0.05\\
 15 keV & 1.888 & 4.24 & 0.05 & 0.42 & 0.05\\
 10 keV & 1.887 & 4.63 & 0.08 & 0.42 & 0.05\\
 7 keV & 1.811 & 4.88 & 0.08 & 0.40 & 0.04\\
 6 keV & 1.713 & 4.80 & 0.07 & 0.40 & 0.03\\
 5 keV & 1.507 & 4.30 & 0.06 & 0.39 & 0.03\\
 4.5 keV & 1.330 & 3.93 & 0.06 & 0.39 & 0.03\\
 4 keV & 1.074 & 3.16 & 0.06 & 0.41 & 0.05\\
\hline
\hline
\multicolumn{6}{c}{\textit{Halo A - Medium Box}}\\
\hline
\hline
 CDM & 18.533 & 4.09 & 0.06 & 0.39 & 0.08\\
 5 keV & 18.894 & 4.63 & 0.06 & 0.41 & 0.05\\
 3 keV & 16.912 & 4.76 & 0.07 & 0.40 & 0.04\\
 2 keV & 10.126 & 3.10 & 0.05 & 0.40 & 0.05\\
\hline
\hline
\multicolumn{6}{c}{\textit{Halo A - Large Box}}\\
\hline
\hline
 CDM & 600.497 & 3.34 & 0.04 & 0.39 & 0.13\\
 2 keV & 604.681 & 4.07 & 0.08 & 0.40 & 0.05\\
 1 keV & 547.367 & 4.44 & 0.07 & 0.39 & 0.03\\
 0.75 keV & 351.273 & 3.48 & 0.04 & 0.38 & 0.05\\
\hline
\end{tabular}
\end{center}
\end{table}

\section{Results I - NON-UNIVERSALITY OF PROFILES}\label{results}

We begin by examining the effects of the WDM power spectra on 
the density structure of Halo A in the three boxes and thus the three 
mass scales of the halo. We then examine the kinematics 
and conclude by checking the convergence.
In Section~\ref{secVAR} we examine a larger halo sample to check if 
the results of Halo A are typical for haloes in general 
and we show the features in the profiles are dynamically stable.

\subsection{Density Structure}

\begin{figure*}
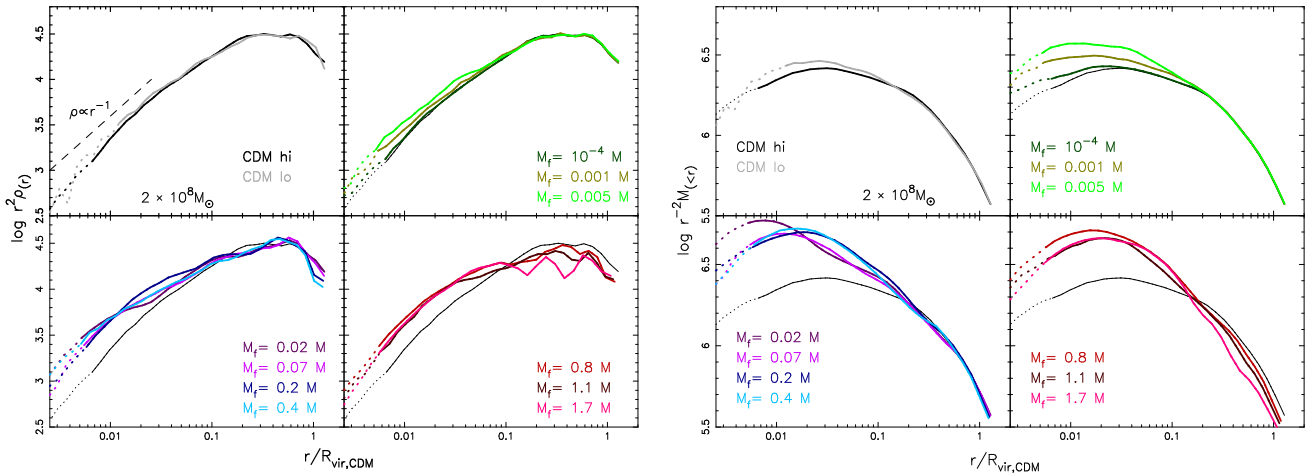

\begin{center}
\begin{tabular}{c c}
\includegraphics[height=1.0\columnwidth,angle=270]{FIG02A.eps} &
\includegraphics[height=1.0\columnwidth,angle=270]{FIG02B.eps} \\
\end{tabular}
\caption{(Left) Density profiles of the small mass simulations of Halo A at the 
normalization time, $a=0.1$. The dashed line gives the asymptotic slope of 
the NFW profile at small radii. (Right) Cumulative mass profiles of the 
small mass simulations of Halo A. The density profiles have been multiplied 
by $r^{2}$ and the mass profiles by $r^{-2}$ to reduce the dynamic range. 
The radial coordinates have been normalized to the virial radius in CDM and 
are plotted to six softening lengths ($6\epsilon$). The WDM profiles have 
been grouped and are plotted against the CDM profile for clarity. All 
profiles are plotted with solid lines where they satisfy the convergence 
criterion of \citet{pow2003}. As the filtering mass increases clear 
deviations from a universal shape are seen in the WDM profiles caused by 
the displacement of mass from intermediate radii to the core in the WDM 
simulations.
\label{figDen3A}}
\end{center}
\end{figure*}

The left panel of Figure~\ref{figDen3A} shows the spherically averaged 
density profiles of Halo A at $a=0.1$ in all cosmologies for the small 
mass simulations, $M=2\times 10^8 M_{\sun}$. The profiles are plotted with 
solid lines where they satisfy the convergence criterion of \citet{pow2003} 
and the inner profiles are extended to six force softening lengths with 
dotted lines. Excellent agreement is seen between the high and low 
resolution CDM profiles. No significant differences from CDM are seen for a 
filtering mass $10^{-4}$ of the halo mass. As the filtering mass increases the 
inner profile at $r < 0.1 R_{vir}$ steepens and the density increases. 
The location where the WDM density begins to increase shows a correlation 
with filtering scale, moving to larger radii as the cosmology gets warmer 
and the filtering scale larger. In the range $0.1-0.4 R_{vir}$ the WDM 
densities fall below that of CDM while the densities agree with CDM for 
$r>0.4R_{vir}$ except in the warmest cosmologies where the halo outskirts 
are still growing by secondary infall. These features are more pronounced 
in the cumulative mass profiles shown in the right panel of 
Figure~\ref{figDen3A}. The enclosed mass is equivalent in CDM and WDM 
at $> 0.5 R_{vir}$ indicating it is the mass in shells at $0.1-0.4 R_{vir}$ 
that has been displaced to smaller radii in the WDM simulations.

\begin{figure*}
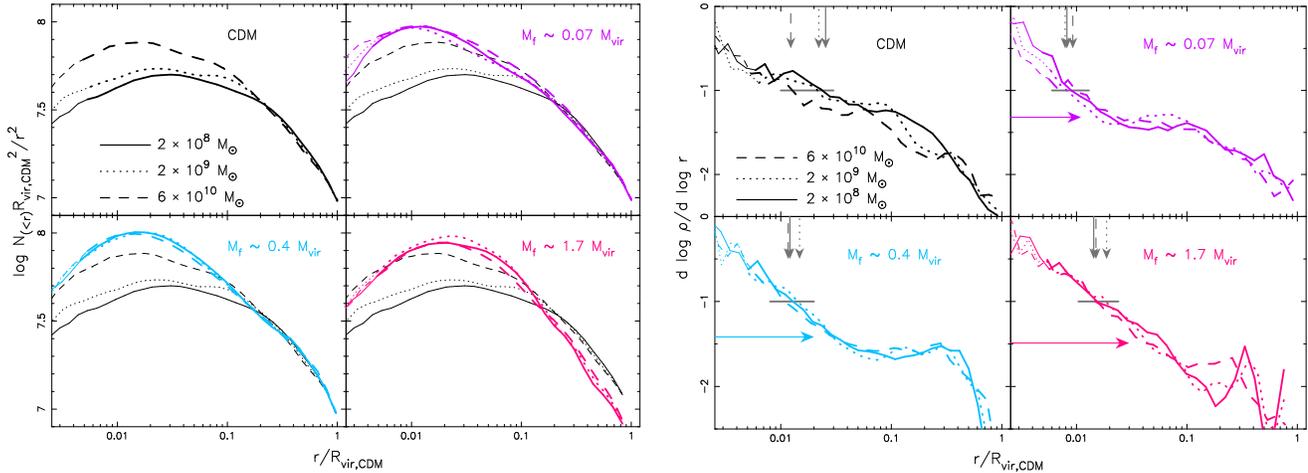

\begin{center}
\begin{tabular}{c c}
\includegraphics[height=1.0\columnwidth,angle=270]{FIG03A.eps} &
\includegraphics[height=1.0\columnwidth,angle=270]{FIG03B.eps} \\
\end{tabular}
\caption{(Left) Comparison of the cumulative mass profiles of Halo A at the 
normalization times of the three mass scales when the haloes have grown to 
$\sim 10^7$ particles. The small, medium, and large mass haloes are plotted 
with the solid, dotted, and dashed lines, respectively. The profiles are 
given by the number of enclosed simulation particles and the radial coordinates 
have been normalized by the CDM virial radii. The WDM profiles are plotted 
against the CDM profiles for comparison.
(Right) Comparison of the slope of the density profiles of Halo A across the 
three mass scales at the normalization times. Short grey lines and vertical 
arrows indicate where the log slope is $-1$ in all cosmologies. The CDM halo 
is steeper at the large mass scale than the small scale. The WDM haloes 
are generally steeper than the CDM but soften at $r<0.01 R_{vir}$.
Horizontal arrows show the fitting formula of \citet{ish2014} and match 
the WDM plateaus to $10\%$.
\label{fignpartpro}}
\end{center}
\end{figure*}

To compare the mass profiles of Halo A across the three mass scales 
we plot in the left panel of Figure~\ref{fignpartpro} the profiles 
of enclosed number of simulation particles and normalize the radial 
coordinates by the CDM virial radius in each box. Interestingly, 
variations are seen at $r<0.1 R_{vir}$ in the CDM haloes as a function 
of the halo mass in contrast to the WDM simulations where the profiles 
are nearly identical across the explored halo mass range. The enclosed 
mass in the CDM inner halo becomes greater as the halo mass increases 
but when small scale structures are erased, as in the WDM simulations, 
the mass profiles are insensitive to the halo mass. Angular momentum 
sets the shape of the inner profile in the models of \citet{del2009} 
where more massive haloes are predicted to have less angular momentum 
resulting in steeper profiles. It can be seen from Table~\ref{tab2} that 
the spin of the CDM halo decreases as the mass scale increases, consistent 
with this idea.

The CDM halo spin parameter is $26\%$ higher at the small mass scale 
compared to the large while the WDM haloes vary by $\lesssim 10\%$
which may be why the WDM profiles are very similar. However, the WDM 
spin parameters are generally higher than the CDM halo at all scales
yet they have steeper profiles than CDM so this is not the entire answer.
One difficulty is the spin parameter includes all particles within the 
virial radius while the greatest differences between profiles are seen 
in the inner halo. This is examined further in Section~\ref{origin}.

It is important to emphasize that the differences between WDM and CDM profiles 
diminish as the halo mass increases due to the steepening of the CDM profile.
This observation may explain why previous works have not clearly identified 
the prominent features and trends in the profile shapes found in this work 
and the recent simulations of microhaloes near the CDM filtering 
scale \citep{ish2010,and2013b,ish2014}.

In the right panel of Figure~\ref{fignpartpro} the logarithmic slope of the 
density profiles are compared across mass scales for a given ratio $M_f/M$. 
The large mass CDM halo profile is steeper than the medium and small haloes 
for $r<0.3 R_{vir}$ and reaches the NFW value of $-1$ at a smaller radius 
(given by the vertical short grey lines). However, the differences are less 
than predicted by the model of undigested subhalo cores of \citet{sub2000} 
but in agreement with the predictions of the angular momentum models 
of \citet{del2010}.

Unlike the CDM profiles the slopes in the WDM cosmologies are nearly identical 
across mass scales. For $M_f < M$ the slopes tend toward plateaus of constant 
value when moving to smaller radii. These plateaus also depend on the filtering 
scale, ranging from $r = 0.02-0.1 R_{vir}$ for $M_f \sim 0.07 M$ but grow steeper 
and move outward to $r = 0.04-0.3 R_{vir}$ for $M_f \sim 0.4 M$. These tendencies 
were also seen in the CDM microhalo simulations of \citet{ish2014}. They fit a 
power law function to the relation between halo mass and inner slope $\alpha$ 
for their sample: 
\begin{equation}\label{eq3_1}
\alpha = 0.123 \log(M_{vir}/M_f) - 1.461,
\\\end{equation} 
with a scatter of $20\%$. We plot the values given by this equation as horizontal 
arrows in the right panel of Figure~\ref{fignpartpro} and find agreement 
with the slope plateaus to $10\%$.

Our simulations also show interesting new features in regimes not explored 
in \citet{ish2014}. Continuing to smaller radii the slopes do not remain at 
the asymptotic values of the plateaus but soften. The slopes remain steeper 
than CDM for $r<0.1 R_{vir}$ and achieve $-1$ at smaller radii, although this 
scale moves outward as the filtering scale gets larger. The inner profiles 
quickly become softer than CDM at $r \lesssim 0.01 R_{vir}$. The inner profiles 
are shallower in WDM in agreement with the models of \citet{wil2004} and 
\citet{vin2012}. For $M_f > M$ the slope profiles do not form a plateau but 
remain steeper than CDM for $r > 0.01R_{vir}$.

\subsection{Internal Kinematics}

\begin{figure}
\begin{center}
\includegraphics[height=0.98\columnwidth,angle=270]{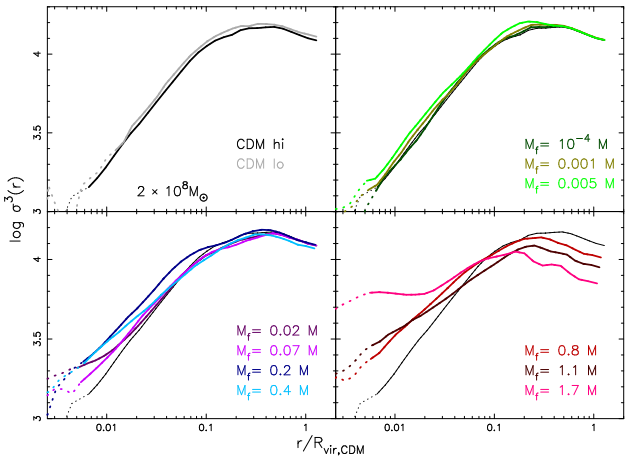}
\caption{Velocity dispersion profiles of the small mass simulations of 
Halo A. WDM simulations have been grouped and plotted against the CDM profile 
for clarity.\label{figsig3}}
\end{center}
\end{figure}

Figure~\ref{figsig3} shows the profiles of $\sigma^3$ for the small 
mass simulations of Halo A, where $\sigma$ is the local 3D velocity dispersion. 
Similar to the density profiles, the dispersions are greater in the 
inner WDM haloes compared to CDM and show a correlation with the 
filtering scale, growing larger and extending to greater radii as the 
cosmology becomes warmer and the filtering scale increases.
This can be understood as a consequence of the increased mass in the WDM cores.
As the mass in the core grows the dispersion must get larger to 
stay in virial equilibrium against the deeper potential well.

\begin{figure*}
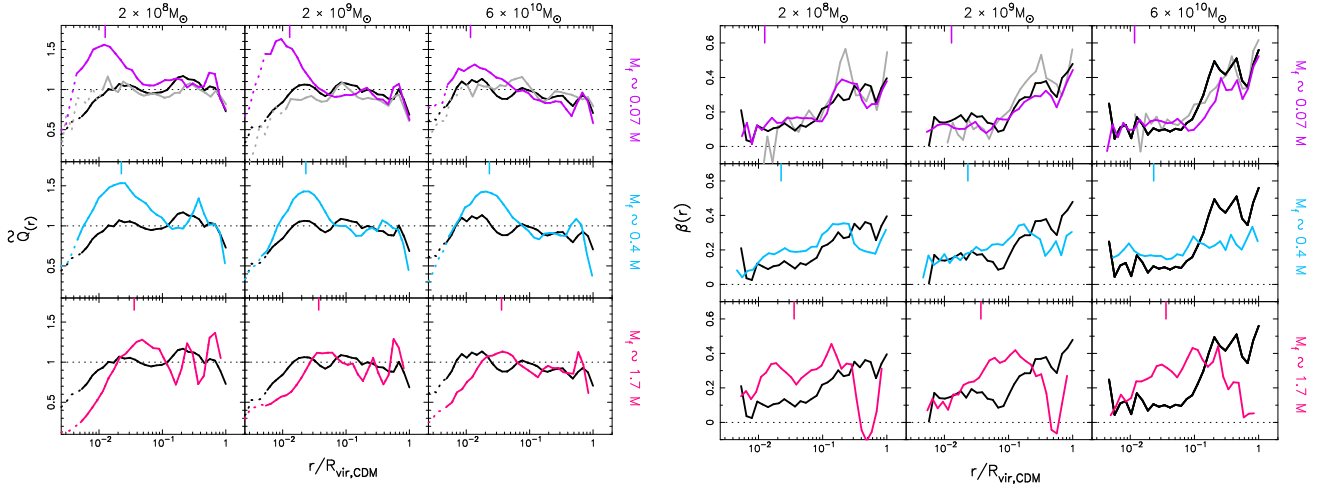

\begin{center}
\begin{tabular}{c c}
\includegraphics[height=1.0\columnwidth,angle=270]{FIG05A.eps} &
\includegraphics[height=1.0\columnwidth,angle=270]{FIG05B.eps} \\
\end{tabular}
\caption{(Left) Deviations from power-law behaviour in the phase space density 
profiles of the simulations of Halo A. (Right) Velocity anisotropy profiles 
of the simulations of Halo A. The simulations are grouped by 
mass scale: small, medium, and large from left to right, and by relation of 
filtering mass to halo mass with the cosmology growing warmer from top to 
bottom. Deviations from power-law are seen in the inner WDM haloes that 
reach a peak at $\sim 4\%$ of the filtering length (coloured ticks along 
the top of each plot). However, no corresponding features are seen in the 
velocity anisotropy. 
\label{figQA}}
\end{center}
\end{figure*}

To examine the phase space density profiles of Halo A we adopt $\gamma=-1.875$ 
and fit the profiles of each CDM halo to the form, $Q_{fit} = A r^{\gamma}$. 
For illustrative purposes we show $Q$ normalized to the power law fit, 
$\tilde{Q}=Q/Q_{fit}$, to emphasize deviations from a simple power law. 
We calculate $\tilde{Q}$ for the WDM simulations using the CDM $Q_{fit}$. 
The left panel of Figure~\ref{figQA} shows the deviations from power 
law for Halo A in common cosmologies for all three mass scales. 
A prominent feature is seen in the inner regions of the WDM haloes, reaching 
a maximum deviation before declining. Along the top axis of each plot are 
ticks marking the location of $0.037 R_f$ in each WDM cosmology. 
This scaling was empirically determined but marks the location of the 
peak remarkably well indicating the deviations scale with the filtering scale.

Interestingly, a drop in the inner profile of the CDM haloes is seen that 
becomes more pronounced as the halo mass decreases. This also
agrees with the models of \citet{del2010} that demonstrate a dependence
of the $Q$ profile on halo mass.

A useful metric of the particle orbits is the velocity anisotropy 
parameter given by:
\begin{equation}
\beta(r) = 1 - \frac{\sigma_\theta^2+\sigma_\phi^2}{2\sigma_r^2},\\
\end{equation}
where $\sigma_\theta^2$ and $\sigma_\phi^2$ are the angular velocity 
dispersions and $\sigma_r^2$ is the radial velocity dispersion.
For purely radial orbits, $\beta=1$, while isotropic particle 
motions give $\beta=0$. At the halo outskirts $\beta \rightarrow 1$ 
where freshly accreted material is still falling inward while in the 
core of a relaxed halo $\beta \rightarrow 0$.
In practice the anisotropy parameter is seldom exactly zero in 
the inner regions since simulated haloes are generally not spherically 
symmetric.

The right panel of Figure~\ref{figQA} shows the velocity anisotropy
profiles of Halo A at common filtering scales in all simulations. There is 
a radial bias in the CDM particle orbits at $r>0.1 R_{vir}$ while particles 
inside this scale are well isotropized. The WDM anisotropy profiles are generally 
similar to CDM, although in the warmest cosmologies (bottom row) the haloes 
have radial bias extending deeper into the inner halo than in the other 
cosmologies. An examination of the halo shapes revealed a tendency for 
the haloes to become more spherical in the inner and outer regions 
at $r<0.02 R_{vir}$ and $r>0.2R_{vir}$ and less spherical in the intermediate 
regions as the filtering mass increases. The features of the velocity 
anisotropy may thus not be due to an incomplete isotropization of the 
particle velocities but simply to the increased triaxiality seen at these radii.
What is clear is the lack of any feature at the location of the peak 
deviations in the $\tilde{Q}$ profile making it apparent the physical 
processes that created the increased mass in the WDM cores do not leave 
an imprint on the isotropy of particle velocities after virialization.

\subsection{Convergence Tests}

\begin{figure}
\begin{center}
\includegraphics[height=0.9\columnwidth,angle=270]{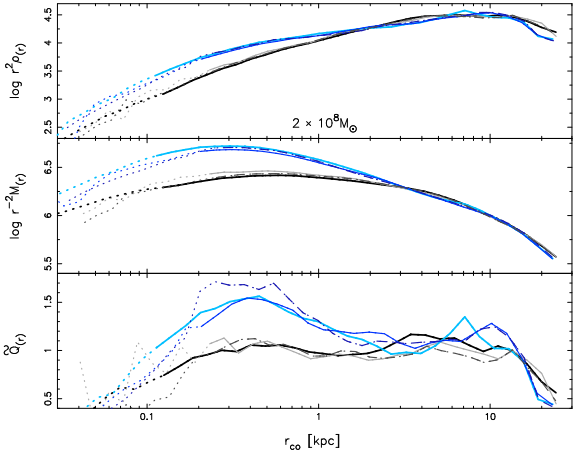}
\caption{Comparison of the high and low resolution small mass simulations of 
Halo A. The high and low resolution CDM simulations are plotted in black 
and grey, respectively, the high and low 6 keV simulations 
in light and dark blue. The dashed lines are the low resolution 
6 keV and CDM simulations started from a higher redshift. Profiles are 
plotted with solid lines where they satisfy the convergence criterion 
of \citet{pow2003} and are extended to $3\epsilon$ with dotted lines. 
Consistent results are seen across simulations demonstrating the 
results are not affected by the mass resolution or starting 
redshift.\label{figConv}}
\end{center}
\end{figure}

Figure~\ref{figConv} shows the profiles of Halo A in the high and low 
resolution simulations and the test simulations initiated from 
a higher redshift. Excellent agreement is seen across resolutions in 
both CDM and WDM and from the simulations started from higher redshift. 
The convergence criterion of \citet{pow2003} tested on CDM simulations 
appears to be not only valid but perhaps overly conservative in measuring the 
convergence radii of our WDM simulations.

Simulations with truncated power spectra are known to produce numerically 
artificial small mass haloes along the filaments of collapsed density 
perturbations \citep{wan2007} whose size and separation are dependent on 
the mass resolution. Figure~\ref{figConv} also shows the results are not 
due to these spurious haloes since a dependence on the mass resolution 
would be expected to reflect on the shape and location of the features 
in the WDM profiles.

\section{Results II - Testing Cosmic Variance}\label{secVAR}

To explore if the results for Halo A were typical of haloes in general we 
simulated two additional haloes, Halo B and Halo C. We first ran a low resolution 
simulation with a cubic refinement volume of side length $1/4$ the box length, 
composed of $512^3$ particles, and centred on Halo A. The 15 largest haloes in 
the refinement volume were examined in detail. Two haloes were chosen based on 
their quiescent accretion histories and relaxation metrics for individual 
resimulation at high resolution in an analogous way to Halo A but only at large 
mass scale and only in CDM, 1.1~keV and 0.75~keV cosmologies. 
Haloes B and C grow to about the same size as Halo A by the end of the simulations but have later formation epochs suggesting Halo A forms from a volume with greater initial overdensity. For example, the 0.75~keV large scale Halo A grows to $N=10^6$ at $a=0.125$ while Haloes B and C don't reach this size until $a=0.148$ and $0.165$, respectively.

We ran the large scale simulations of Haloes B and C to $a=0.25$ to test the 
stability of the profiles over a greater time span than the simulations on 
Halo A presented in Section~\ref{results}. To test the dynamical stability of 
Halo A we ran an extended simulation set at small mass scale for CDM, 6~keV 
and 4~keV WDM. The refinement volume was increased a factor of two to 
allow Halo A to evolve to $a=0.125$ while preventing contamination by accretion 
of particles from outside the refinement volume. Snapshots of the particle 
data were saved every 0.001 change in scale factor.
We list the properties of all three haloes at the end of their simulations in Table~\ref{tab3}.

\begin{table}
\caption{Properties of the small scale Halo A and large scale Haloes B and C at the end of each extended simulation set.\label{tab3}}
\begin{center}
\begin{tabular}{l c c c c c}\hline
 Cosmo & $M_{vir}$ & $\lambda^{\prime}$ & $x_{off}$ & $\frac{2K}{|U|} - 1$ & $f_{sub}$ \\
 & [$10^{8}M_{\odot}$] & [$10^{-2}$] & [$R_{vir}$] & & \\
(1) & (2) & (3) & (4) & (5) & (6) \\
\hline
\hline
\multicolumn{6}{c}{\textit{Halo A -- Small Box -- $a=0.125$}}\\
\hline
\hline
 A CDM & 3.789 & 4.87 & 0.07 & 0.41 & 0.04 \\
 A 6 keV & 4.029 & 4.86 & 0.04 & 0.40 & 0.01 \\
 A 4 keV & 3.362 & 3.25 & 0.04 & 0.37 & 0.01 \\
\hline
\hline
\multicolumn{6}{c}{\textit{Haloes B \& C -- Large Box -- $a=0.25$}}\\
\hline
\hline
 B CDM & 802.734 & 1.96 & 0.03 & 0.24 & 0.09 \\
 B 1.1 keV & 708.649 & 1.77 & 0.02 & 0.21 & 0.01 \\
 B 0.75 keV & 565.010 & 1.23 & 0.04 & 0.21 & 0.01 \\
\hline
 C CDM & 572.440 & 2.72 & 0.05 & 0.21 & 0.12 \\
 C 1.1 keV & 451.046 & 2.78 & 0.06 & 0.19 & 0.02 \\
 C 0.75 keV & 308.477 & 1.45 & 0.02 & 0.17 & 0.01 \\
\hline
\end{tabular}
\end{center}
\end{table}

\begin{figure}
\begin{center}
\begin{tabular}{c c}
\includegraphics[height=0.98\columnwidth,angle=270]{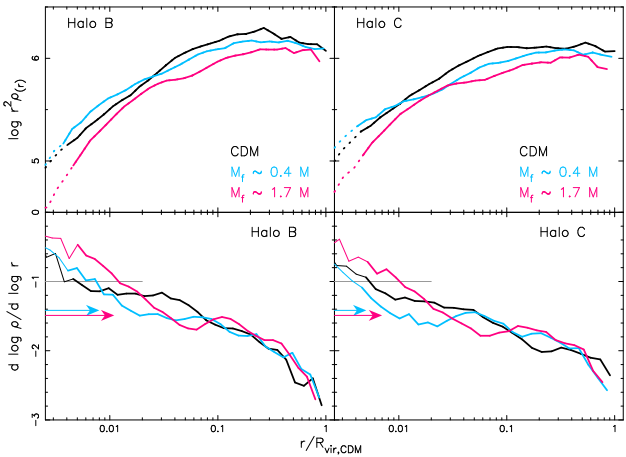}
\end{tabular}
\caption{(Top) Density profiles of Haloes B and C in the large box-size simulations. The density profiles have been multiplied by $r^{2}$ to reduce the dynamic range. (Bottom) Slope of the density profiles for the same two halos.
\label{figDenBC}}
\end{center}
\end{figure}

Figure~\ref{figDenBC} shows the density profiles and slopes for Haloes B and C and give excellent agreement with the results from Halo A. The slopes of the WDM profiles show the same tendency to steepen, reach a plateau, then soften in the inner halo, and show the same dependence on filtering scale. The density profiles similarly show the same characteristic feature of mass displaced from intermediate regions to the core in the WDM cosmologies. The main difference is a stronger deviation from CDM with a larger overall reduction in densities. The density profiles are everywhere below the corresponding CDM halo for the 0.75~keV cosmologies. Halo C shows a larger deviation from CDM than Halo B, likely related to the later formation epochs of these haloes.

\begin{figure*}
\begin{center}
\includegraphics*[width=0.95\textwidth,angle=0]{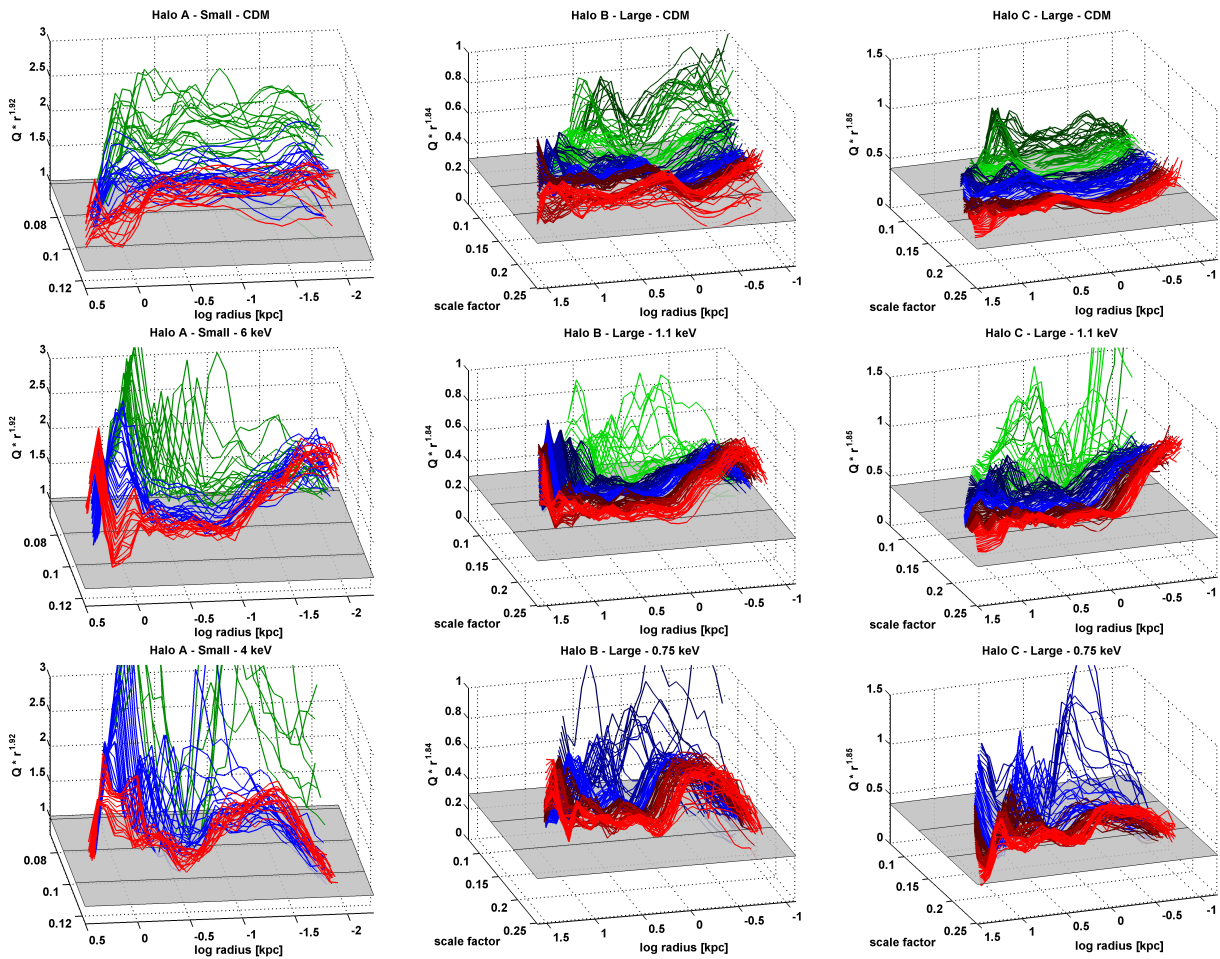}
\caption{Evolution of the phase space density profiles of Halo A at small scale 
and Haloes B and C at large scale (left to right). CDM cosmologies are plotted 
in the top row, WDM in the bottom rows with filtering scale increasing top to bottom. 
The phase space density profiles 
have been multiplied by $r^\gamma$ to show deviations from power law behaviour. 
The CDM profiles are consistent with power laws with minor fluctuations while 
the WDM profiles show prominent features in their cores that form shortly after 
the growth rate slows and are stable thereafter. Profiles are plotted 
with variable colour simply for visual clarity.
\label{figQ3}}
\end{center}
\end{figure*}

We show the normalized phase space density profiles, $\tilde{Q}$, in 
Figure~\ref{figQ3} for all three haloes in their CDM and WDM simulations 
at times when the haloes are composed of $> 10^5$ particles. The values 
of $\gamma$ for $Q_{fit}$ were determined separately for each halo by 
fitting the CDM profiles at late times. The CDM profiles of all three 
haloes are consistent with minor fluctuations about power laws while the 
large deviation seen in the WDM core of Halo A and its dependence on the 
filtering scale is also seen in the other two haloes.

The prominent feature of the 6~keV WDM core first appears in the profile of small 
scale Halo A at $a = 0.075$, and at $a = 0.115$ in the large scale 1.1~keV 
Haloes B and C. After its formation the core feature is stable through the end 
of the simulation, spanning a time range of 420 Myr for Halo A and 1520 Myr 
for B and C. The dynamical time, $t_{dyn}=\sqrt{R^3/GM}$, for the core of Halo A is 
6.5~Myr and 14-18 Myr for the cores of Haloes B and C, demonstrating the profiles 
are stable over at least $65-85$ dynamical times. 
In the warmer cosmologies the cores form later but are also stable through the end of the simulations.

The mass growth rates, $\dot{M} \equiv d \log M/d \log a$, when the core features 
appear in the WDM profiles are $\dot{M} = 6-10$. At earlier times the growth 
rates are higher and the profiles fluctuate wildly. The fast and slow growth 
phases of \citet{zha2003a} correspond to growth rates in the matter dominated era 
of $\dot{M}=6$ and $\dot{M}=1.5$, respectively. Our result that the core features 
appear as the growth rate slows supports this inside-out view of halo growth.

We conclude the structural and dynamical features of Halo A are valid 
for haloes near the filtering scale of WDM cosmologies in general
and form as haloes transition from fast growth to the slow accretion phase,
remaining stable thereafter.

\section{Origin of the Core}\label{origin}

The simulations in this work have shown changes in the density and 
dynamical profiles in the core is common for haloes near the filtering 
mass in WDM cosmologies. The previous section determined the formation 
time of the core and demonstrated its stability with time. In this 
section we investigate clues to the core's origin.

\begin{figure*}
\begin{center}
\includegraphics[height=0.8\textwidth,angle=270]{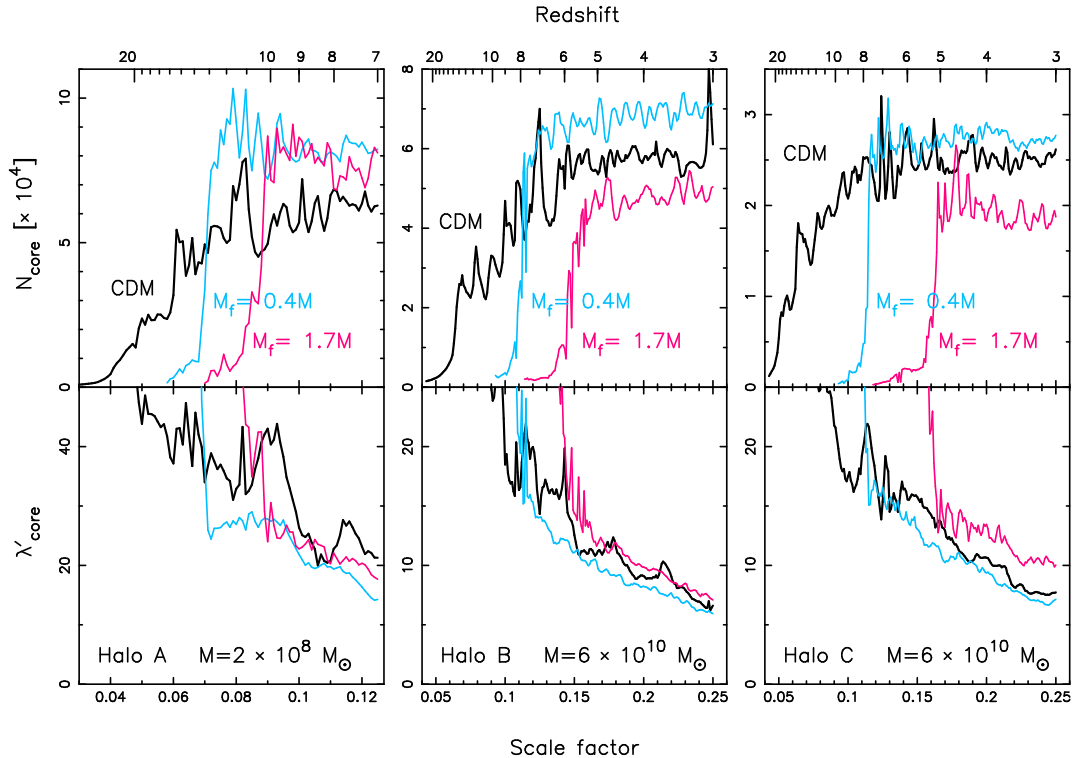}
\caption{Evolution of the number of particles within the core (top) and 
the spin parameter of the core (bottom) in the extended small mass scale 
simulations of Halo A and large mass scale simulations of Haloes B and C 
(left to right). The core radius was held fixed in proper length at 60 pc 
for Halo A, 680 pc for Halo B, and 400 pc for Halo C.  
\label{figCoreA}}
\end{center}
\end{figure*}

We examine the cores of Halo A in the extended small scale simulations 
and the cores of Halo B and C at large mass scale. We define the proper 
radius of the core as 60~pc for Halo~A, 680~pc for Halo B, and 400~pc 
for Halo C. These radii were held constant in proper length and the 
number of particles within each core was calculated for all cosmologies 
at all times a halo progenitor was identified. The top row of 
Figure~\ref{figCoreA} shows the growth in core particles in the CDM and 
WDM simulations. It is clear the core forms quickly in WDM while the 
core is built-up more gradually in CDM. The number of core particles 
remain approximately constant after formation in both CDM and WDM. The 
epoch of WDM core formation occurs shortly after the halo virial mass 
catches up to the CDM halo and the growth rate slows to the CDM rate. 
For example, in the 6~keV small mass scale simulation of Halo A the 
core forms at $a \sim 0.075$, while Figure~\ref{figMASSA} shows the 
growth rate slows to approximately the CDM rate at about the same time. 
This agrees with the formation epoch of the core feature in 
the $\tilde{Q}$ profiles.

We characterize the angular momentum of each core with the dimensionless 
spin parameter, Eqn.~\ref{lamb}, and compare the evolution of each core 
spin in the bottom row of Figure~\ref{figCoreA}. 
There is a clear trend between the number of core particles and core spin. All cores tend to lose angular momentum over time but cores with fewer particles consistently have higher spin. There is also a trend that as the filtering scale increases the number of core particles first increases then decreases. This is most evident in Haloes B and C and is again likely due to the later collapse epochs of these haloes.

\begin{figure}
\begin{center}
\includegraphics*[width=\columnwidth,angle=0]{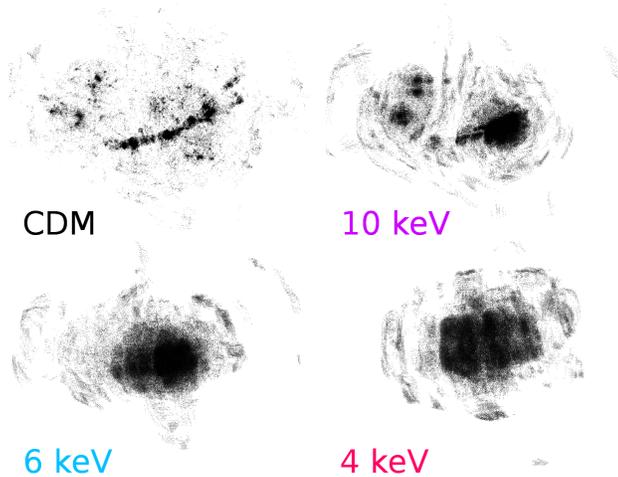}
\caption{Positions at $a=0.05$ of core particles in select cosmologies of the 
small mass scale simulations of Halo A. Images are centred on the centre of 
mass of core particles.\label{figZ19}}
\end{center}
\end{figure}

In Figure~\ref{figZ19} we show the core particles of Halo A in the small 
mass scale simulations at $a=0.05$, well before core formation. The images 
are centred on the particles' centre of mass. As noted by \citet{bus2007}, 
the filtered power spectra cause what were multiple clumps in CDM to become 
one collapsing clump in WDM. The core particles additionally become more 
symmetrically distributed around the centre of mass as the cosmology 
becomes warmer.

This evidence argues for angular momentum playing an important role 
in determining the structure of the core. The importance of angular 
momentum for the shape of the inner profile has been emphasized by a 
number of studies \citep{hus1999a,hus1999b,hio2002,asc2004,lu2006}.
Purely radial orbits give a steep inner profile, $\rho \propto r^{-2.25}$ 
\citep{ber1985}. As the amount of angular momentum is increased particles 
remain closer to their maximum orbital radii resulting in shallower density 
profiles. Angular momentum is dominated by the tangential component of the 
velocity dispersions which are acquired dynamically in both the CDM and 
WDM simulations since thermal velocities were not added to the WDM particles.
Interactions with substructure and the global tidal field produce tangential 
components to the particle velocities. An alternative possibility is radial 
orbit instability \citep{bel2008}. A detailed study of particle orbits is 
outside the scope of this paper, however,
after particles collapse the virialization process isotropizes their velocities 
equally well in both CDM and WDM as seen from Figure~\ref{figQA}.
The higher accretion rates in the WDM fast growth phase may also play a 
role in generating the core as seen in the models of \citet{lu2006}.

\section{Discussion}\label{disc}

We found the inner structures of dark matter haloes in cosmologies 
with truncated power spectra deviate from their profiles in 
non-truncated cosmologies with mass displaced from the intermediate 
regions to the centre. 
We have already shown how our results are in agreement with previous work on CDM 
microhaloes \citep{ish2010,and2013b,ish2014}, 
in this section we discuss how our work compares to other WDM and HDM studies.

\begin{figure}
\begin{center}
\includegraphics[height=0.98\columnwidth,angle=270]{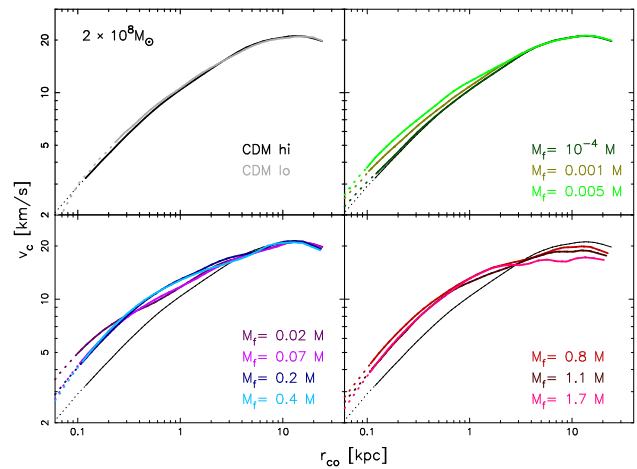}
\caption{Circular velocity profiles of the small mass scale simulations 
of Halo A. The WDM profiles are grouped and plotted against the 
CDM profile for clarity.
\label{figvc7}}
\end{center}
\end{figure}

Investigations of Milky Way satellites in $1-4$~keV WDM cosmologies
have shown the maximum circular velocity decreases and the radius where 
this occurs increases for dwarf galaxy-sized haloes 
\citep{lov2012,and2013,pol2014}. Figure~\ref{figvc7} shows the circular 
velocity profiles of the small mass scale simulations 
of Halo A at the normalization time. It is clear the rearrangement of mass 
in the inner regions has not affected the maximum circular velocity 
or its location to be in disagreement with the conclusions of previous works.

It is well established that in WDM cosmologies haloes below the truncation 
scale form later and have lower concentrations than CDM haloes of similar 
size \citep{avi2001,bod2001,kne2002}. 
In contrast, \citet{and2013b} examined their three CDM microhaloes and found higher concentrations when NFW profiles were fit to the haloes in the filtered power spectrum simulations. However, they found fitting generalized profiles, where the inner slope is allowed to be a free parameter, shifted the scale radius $r_s$ outward and reduced the concentrations.

We investigated the halo concentrations in the low resolution, large refinement volume simulations described in Section~\ref{secVAR}. These simulations were run for the small and large box sizes and 6~keV and 1.1~keV cosmologies, respectively. We fit NFW and generalized density profiles to the nine largest haloes in the refinement volume that were composed of $>10^5$ particles and satisfied the relaxation metrics at the normalization times. We found seven of the nine haloes had greater NFW concentrations in WDM than CDM. However, six of the nine halos had lower concentrations when generalized profiles were fit, consistent across box sizes and consistent with the microhalo results \citep{and2013b,ish2014}. We found the same results when comparing the concentrations of Haloes A, B, and C at the end of the high-resolution extended simulation set. Interestingly, in the simulations with $M_f \sim 1.7 M$ the NFW concentrations of Haloes B and C were less than their CDM concentrations while Halo A was greater. This is consistent with the density profiles in Figure~\ref{figDenBC} and the later formation epochs of Haloes B and C, that enhance the filtering of substructure for a given filtering scale.

Our WDM halo profiles are similar to the profiles seen by \citet{avi2001} and \citet{col2008}.
\citet{col2008} simulated five galactic-sized haloes in WDM and found profiles were 
steeper and denser in the inner region than the best-fitting NFW profiles. 
Additionally, our simulations explain a feature seen in the HDM cluster 
simulations of \citet{wan2009}. The stacked phase space density profiles 
of their 20 most massive haloes show a flattening in the inner 
$0.05 R_{vir}$ in HDM similar to the features seen in the warmest 
simulations of Figure~\ref{figQ3}.

Our results seem to conflict with the work of \citet{bus2007}.
They evolved their CDM and WDM simulations into the future until the
scale factor $a=100$. Past the current epoch the cosmological constant
quickly dominates the density of the universe ($\Omega_\Lambda
\rightarrow 1$) leading to exponential expansion and the supression of
structure growth at $a \sim 3$.  Thus, examining halo properties in
the far future guarantees the haloes have ample time to relax into
their equilibrium states. They find the average density profiles for
haloes near the filtering mass are well fit by the NFW profile for
$r>0.05R_{vir}$ with only lower concentrations below the filtering
mass (although the outer slope in both CDM and WDM is steeper due to
the inflating universe, as noted by \citealt{ric2003}). However, there
are two things that complicate comparison of these simulations to
ours. First is the difficulty due to the different epochs the haloes
are examined at. We examine our haloes shortly after the end of the
fast accretion phase when the inner profile is set but the outskirts
are still growing while \citeauthor{bus2007} examine their haloes well
after all halo growth has stopped and $R_{vir}$ has reached a
maximum. The effects we see in the inner halo will therefore be at
radii smaller than the convergence radius in the haloes simulated by
\citeauthor{bus2007}. Also, they explored haloes at a larger mass
scale and used a much greater filtering mass, $1.2\times 10^{14}
M_{\sun}$. In this work we have shown how even the density profiles of
CDM haloes have a dependence on mass with smaller differences between
CDM and WDM profiles for larger halo masses, for a fixed ratio of the
filtering mass to halo mass.

Finally, we comment on the effects of adding thermal velocities appropriate to
the adopted WDM models to the simulation particles. Thermal WDM particles 
decouple with a finite fine-grain phase space density that imposes an upper 
limit on their density, resulting in soft cores in collapsed haloes.
The radius of this core depends on the mass of the WDM particle and the mass 
of the halo \citep{hog2000}. For the warmest cosmologies of Halo A the 
core radius is $\sim 4 \times 10^{-4} R_{vir}$ which agrees 
with the core sizes seen in the simulations of \citet{mac2012}. 
The thermal core would be about the size of the adopted softening lengths, 
far below the scales where the WDM profiles deviate 
from the CDM profile.

\section{Summary}\label{summary}

We tested the claim that the virialization process erases all 
information about the initial conditions and produces universal 
profiles in gravitationally collapsed dark matter haloes. We simulated 
an isolated halo with an early formation epoch at three mass scales 
from $2 \times 10^8 M_{\odot}$ to $6 \times 10^{10} M_{\odot}$ in CDM and a 
variety of WDM cosmologies where the formation of structures below the 
filtering scale is suppressed. We examined when the halo 
was composed of $\sim 10^7$ particles at each scale and the halo was 
in the slow growth phase. 
We studied two additional haloes to account for cosmic variance. 
We found the haloes were changed both structurally and 
dynamically by the truncation in the WDM power spectra. 
Substructures in the mass range $10^{5-6} M_{\odot}$ have a detectable effect on the slope of the inner density profile in $10^{8-9} M_{\odot}$ haloes because when substructure on this mass scale is filtered out in the 34~keV WDM cosmology the profile visibly deviates from CDM, becoming steeper.
We summarize our main findings below.

\begin{itemize}

\item Density profiles are steeper in WDM than CDM for haloes near the 
filtering scale. For $M_f < M$, the slopes approach a constant value moving 
from the virial radius to the inner halo, whose value is well described by 
the fitting formula of \citet{ish2014}. The value of the constant slope and 
its radial span depend on the filtering scale, moving outward and becoming 
steeper as the ratio of the filtering to halo mass gets larger. Our 
simulations also examined the regime $M_f > M$ and found the density profiles 
do not approach a constant asymptotic slope at intermediate radii but continue 
to flatten monotonically. At smaller radii, for all the masses and WDM 
cosmologies, the slopes do not remain at the constant value found at 
intermediate radii but grow shallower and become softer than CDM for 
$r<0.01 R_{vir}$, in agreement with halo models \citep{wil2004,vin2012}.

\item Particle velocity dispersions increase in the inner profiles while 
velocity anisotropies after virialization are largely similar across cosmologies. 
The changes in density and velocity dispersion create a deviation from a simple 
power law in the inner phase space density profiles. This deviation reaches a 
maximum at a radius proportional to the filtering scale.

\item The core features of the profiles are set early when the halo mass growth 
rate $\dot{M} \sim 6-10$. At higher growth rates the profiles exhibit large 
fluctuations. After formation the core features are dynamically stable.

\item The halo mass structure is rearranged in WDM compared to CDM with radii 
$< 0.1 R_{vir}$ gaining mass at the expense of radii $0.1-0.4 R_{vir}$. Furthermore, 
the core is built up gradually in CDM from particles distributed asymmetrically 
in clumps about the centre of mass in contrast to WDM where the core is formed 
in an impulsive event from particles distributed smoothly and symmetrically. 
There is a correlation between the number of particles and the spin of the core 
with more massive cores having less spin. 
This argues for angular momentum as the physical mechanism responsible for the 
differences in profiles as seen in the models of \citet{del2009,del2010}.

\item We found a dependence on mass in the CDM profiles with larger haloes 
exhibiting a steeper density profile as in \citet{ric2003}. The spin parameter 
decreases with increasing mass in agreement with the models of \citet{del2009} 
that more massive haloes have less angular momentum resulting in steeper 
profiles. However, the effects are much weaker than those of the truncated power 
spectrum. The WDM haloes had similar spins across mass scales and 
also had similar profiles.

\end{itemize}

Our work complements and reinforces the results of recent simulations of 
microhaloes near the CDM filtering scale \citep{ish2010,and2013b,ish2014} and 
shows that the shape of halo profiles cannot be parameterized 
simply by a generalized NFW or Einasto profile with a concentration 
or scale radius dependent on the mass or cosmology. The halo shape is more 
complex, with logarithmic slopes that can vary non-monotonically 
and with features in the profile that reflect the shape of the initial matter power spectrum. 
Thus, in general haloes cannot be fitted by a universal density profile. 
This is actually good news because it may become feasibile to find 
fingerprints of the initial power spectrum of perturbations on galactic 
or sub-galactic scales in the density profiles of dark matter dominated 
dwarf galaxies or clusters.

\section*{Acknowledgments}
The simulations presented in this work were run on the 
{\sevensize DEEPTHOUGHT} computing cluster
at the University of Maryland College Park, the Cray XE6 {\sevensize GARNET} 
at the U.S. Army Engineer Research and Development Center and 
the SGI Ice X {\sevensize SPIRIT} at the U.S. Air Force Research Laboratory.
Basic research in astrophysics at NRL is funded by the U.S. Office of 
Naval Research. EP acknowledges support under the Edison Memorial Graduate
Training Program at the Naval Research Laboratory. 
The authors acknowledge the University of Maryland supercomputing resources 
(http://www.it.umd.edu/hpcc) made available in conducting the
research reported in this paper. MR's thanks the National Science Foundation and NASA for support under the grants NASA NNX10AH10G, NSF CMMI1125285 and the Theoretical and Computational Astrophysics Network (TCAN) grant AST1333514.

\bibliographystyle{mn2e}
\bibliography{Fingerprints}

\label{lastpage}

\end{document}